# Provide Proactive Reproducible Analysis Transparency with Every Publication

Paul Meijer, Nicole Howard, Jessica Liang, Autumn Kelsey, Sathya Subramanian, Ed Johnson, Paul Mariz, James Harvey, Madeline Ambrose, Vitalii Tereshchenko, Aldan Beaubien, Neelima Inala, Yousef Aggoune, Stark Pister, Anne Vetto, Melissa Kinsey, Tom Bumol, Ananda Goldrath, Xiaojun Li, Troy Torgerson, Peter Skene, Lauren Okada, Christian La France, Zach Thomson, Lucas Graybuck

Allen Institute for Immunology

## Abstract

The high incidence of irreproducible research has led to urgent appeals for transparency and equitable practices in open science. For the scientific disciplines that rely on computationally intensive analyses of large data sets, a granular understanding of the analysis methodology is an essential component of reproducibility. This paper discusses the guiding principles of a computational reproducibility framework that enables a scientist to proactively generate a complete reproducible trace as analysis unfolds, and share data, methods and executable tools as part of a scientific publication, allowing other researchers to verify results and easily re-execute the steps of the scientific investigation.

Reproducibility became a focus of debate in the scientific community more than a decade ago, when Amgen researchers revealed that of 53 landmark preclinical cancer studies analyzed, the results of only 6 (11%) could be reproduced (Begley & Ellis, 2012). In the field of psychology, a similar effort showed that results could not be replicated in ⅓ to ½ of the 100 studies analyzed (Open Science Collaboration, 2015). A survey of 1500 scientists in chemistry, physics, engineering, earth and environmental sciences, biology and medicine found that, on average across these disciplines, more than 70% failed to reproduce their own or other scientists' findings (Baker, 2016a). These and similar studies show that the inability to reproduce published results is common and not limited to a particular field (Harris, 2018).

Multiple factors are thought to contribute to this lack of reproducibility, including poor experimental design, lack of (or poor adherence to) standard operating procedures in the experimental lab, poor data analysis, improper interpretation of statistics, unavailability of the original data, selective reporting, the omission of key details about an experiment in published work, shortcomings in the peer review process, a tendency to undervalue studies that fail to reproduce or replicate, the pressure to publish, creative license and outright fraud, as well as other societal and career pressures leading to misconduct (Baker, 2016a; Begley, 2013; Begley & Ellis, 2012; Harris, 2018; Sarewitz, 2016; Stroebe et al., 2012; Van Noorden, 2023; Zhang et al., 2022). In response to these findings, the scientific community has widely endorsed open science practices, such as greater transparency in design and analysis and increased sharing of data for examination. However, even with improved transparency, reproducibility remains difficult in many cases due to incomplete data sets and missing steps in the analysis (Hardwicke et al., 2021; Kidwell et al., 2016).

In the current study, we focus on reproducibility as it pertains to data analysis, reporting and dissemination of results in life sciences. We describe the guiding principles of an open scientific computing framework that empowers research teams to document their data analysis in real time. After publication, the framework supplies to the scientific community both the data and the computational tools necessary to verify the analysis undertaken, enabling outside scientists to execute the same computations as originally performed. By furnishing the necessary infrastructure for scientific computing, the framework eliminates the requirement for external scientists to establish their computing environment, thus lowering the barrier to entry and ensuring fair access.

We assert that fostering proactive data traces is crucial for promoting analysis reproducibility. Institutional policies ought to endorse these practices in research analysis by rewarding researchers who adhere to them. As Big Data and intense computation become commonplace in science, equitable access to data, methods and computing infrastructure helps ensure the participation of the entire scientific community.

# Failures versus Flaws

The failure to replicate a study's findings could stem from errors in the original research design or execution. Alternatively, seemingly minor variations introduced during the replication attempt might inadvertently introduce confounding variables or reveal alternative explanations for the observed effects. The analysis framework outlined here can help reveal either possibility. Let's examine both scenarios.

Reproducibility is a multifaceted concept, and an inability to reproduce a particular study may hint at scientifically relevant—albeit perhaps nuanced—differences between the original and the replicated study. If we use the term "replication" to refer to rerunning an entire study, for instance by redoing the entire experiment with new subjects or materials and carefully reproducing the steps of a previous study, then a failure to replicate the original study's findings may help expose confounding variables and further the scientific debate. In this version of reproducibility, the scientific debate unfolds as studies build on each other—a form of reproducibility essential to scientific progress (Oreskes, 2021; Redish et al., 2018). This phenomenon has been described as a scientific failure (Firestein, 2015).

Alternatively, the inability to reproduce study results may be caused by errors or mistakes unwittingly introduced during execution of either the original research or the reproducibility attempt. This phenomenon, which has been described as a research flaw, detracts from the larger scientific debate (Firestein, 2015).

Our computational reproducibility framework follows guiding principles intended to clarify the analytic phase of an experiment by achieving computational reproducibility (Baker, 2016b). This set of guiding principles makes it easier to detect scientific flaws both before publication and in published work. It also helps reveal scientific failures by delineating the analytical approach and facilitating a debate on alternative analytic approaches that may produce different outcomes. This study is not directly concerned with the wet lab, that is, the data generation processes in the research lab prior to analysis, although we believe that standardizing data generation procedures is a key prerequisite for standardizing analysis.

Next we turn to describing the guiding principles of our computational framework, which are "Ensuring Proactive versus Retroactive Transparency", "Tracking Data and Transformations", "Streamlining Administrative Overhead through Automation", "Advertising Transparency", "Using Executable Tooling", and "Supporting Open Science with Equitable Access". We then demonstrate how these principles translate into practice by applying the framework to analyze data from various immunology studies.

# Proactive versus Retroactive Transparency

While scientific journals and granting agencies like the National Institutes of Health (NIH) strongly advocate for transparency in data and methods, they are less outspoken about the processes that enable such openness (Lowenberg & Puebla, 2022; National Institutes of Health, cited Jan 2023). A key question regarding achieving research transparency is when these efforts should begin. One approach, often seen as the default, is to address transparency only after research completion and manuscript preparation. This is often because data dissemination might be a requirement for publication. Let's call this the retroactive method. In contrast, a proactive approach advocates for continuous tracking of data and methods throughout the research process. Our computational framework champions the proactive approach.

In a retroactive approach, a scientific team retraces its steps to reconstruct its analysis path. If this backtracking process occurs around the time of manuscript submission, there can be a substantial delay between the actual analysis and its retroactive recording. This delay increases the likelihood of missing key details.

In a proactive approach, the scientific team actively tracks analysis methodology as the study unfolds. This real-time approach creates a more reliable trace of the analysis process. Furthermore, automating trace construction within the framework further enhances reliability. We'll discuss the importance of reducing administrative overhead through automation later. Ultimately, by publication time, the framework holds a complete trace that can be readily shared with external scientists for review and replication.

The ongoing push for open science practices is a positive step towards research transparency (Christensen et al., 2020).. However, simply sharing more data and tools doesn't guarantee completeness. Current practices may result in partially reconstructed analysis traces, making it difficult for other researchers to find, understand, and ultimately reproduce the original findings (Read et al., 2015; Roche et al., 2015). This lack of complete information can limit the ability to verify the research and potentially hinders scientific progress (Asendorpf et al., 2013; Hardwicke et al., 2021; Kidwell et al., 2016). While discrepancies in analysis might not always materially impact the conclusions (Hardwicke et al., 2018, 2021), championing a proactive approach to obtain a complete trace of the analysis can significantly enhance the trustworthiness and reproducibility of research.

The proactive method focuses not only on achieving openness and reproducibility of published material for external scientists, but also on extending this transparency to peers in the laboratory as the research is undertaken. This shift allows for earlier, continuous verification of results, greatly enhancing the likelihood that the eventual published research will include a complete, usable set of data and methods (Lowenberg & Puebla, 2022; Martone & Nakamura, 2022a, 2022b).

# Tracking Data and Transformations

To ensure a complete audit trail, both the data and the data manipulations must be tracked. This includes capturing raw data, quality control steps, primary analyses (e.g. per-sample or per-subject), filtering decisions, secondary analyses (e.g., cross-sample or cross-subject), visualizations, and any other relevant transformations. By capturing this detailed record, researchers gain complete insight into the exact analytic procedures employed which is imperative for transparency (Hardwicke et al., 2021), and allows researchers to evaluate the chosen analysis strategy against valid alternative approaches (Botvinik-Nezer et al., 2020; Gould et al., 2023; Silberzahn et al., 2018).

An example of tracking data and transformation is shown in Figure 1. It shows the data ingestion of multiple files, which were then taken through a QC process and combined into a single FCS file. Next a gating process produced a file containing statistical results. This file was used by an analyst in a coding environment to produce a visualization.

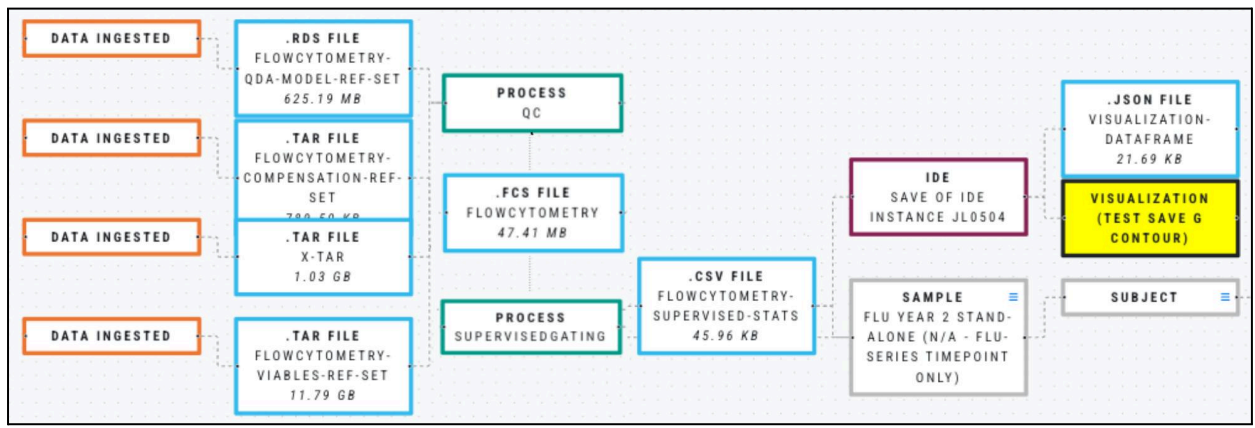

Figure 1. The full provenance of an analysis.

Note that any modification generates a new data file, which is stored separately. Tracking each data modification as a discrete entity preserves the data at each step, ensuring the integrity of the trace for reproducibility studies.

To ensure the reproducibility of data transformations, it's essential to meticulously record the full execution details of each step (Heil et al., 2021; Perkel, 2023). This comprehensive record should encompass:
- *Code and Configuration*. The exact code used for the transformation, along with any configuration files or settings that define its behavior.
- *Dependent Libraries*. A list of all software libraries the code relies on, including their specific versions.
- *Runtime Environment*. A description of the computational environment where the code was executed, including the operating system, hardware specifications, and any relevant software versions.
- *Data Dependencies*. Information about the reference datasets or trained models used in the transformation step.
- *Additional Components*. Any other elements crucial for replicating the computational environment, such as custom scripts or tools.

Detailed metadata associated with the subject or sample, the data itself and the transformation process further enhances the interpretability of the trace. Subject and sample metadata describing the (non)human subjects and specimens being studied, must be included in the full dataset as this will affect the interpretation of the results (Harris, 2018). Examples of such metadata include subject demographics and specifics of sample collection.

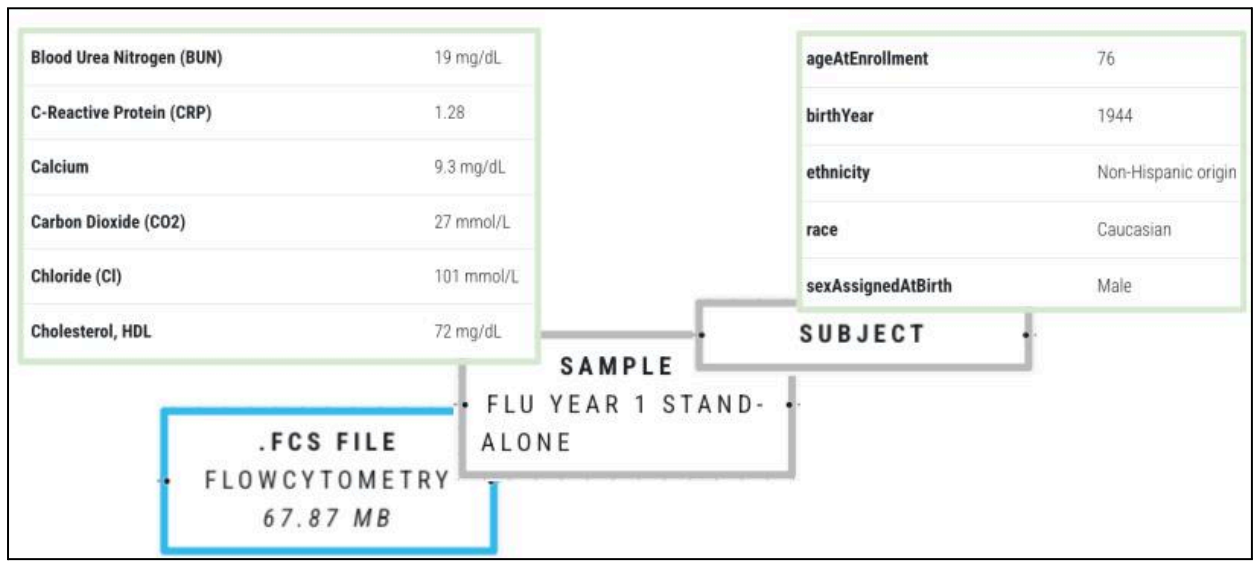

Figure 2. An example of subject metadata associated with a data set.

Metadata on the data itself may include creation date, storage information (e.g., location and file type), data format (e.g., CSV or H5), and data modality (e.g., microscopy image, single cell RNA sequences, or reaction time data). This comprehensive information helps researchers understand the origin, format, and nature of the data, ultimately facilitating accurate analysis and interpretation of the results.

Finally, structural metadata on the transformation process, including input data, transformation type, and output data, creates a connected trace by providing the critical links between data and its transformations.

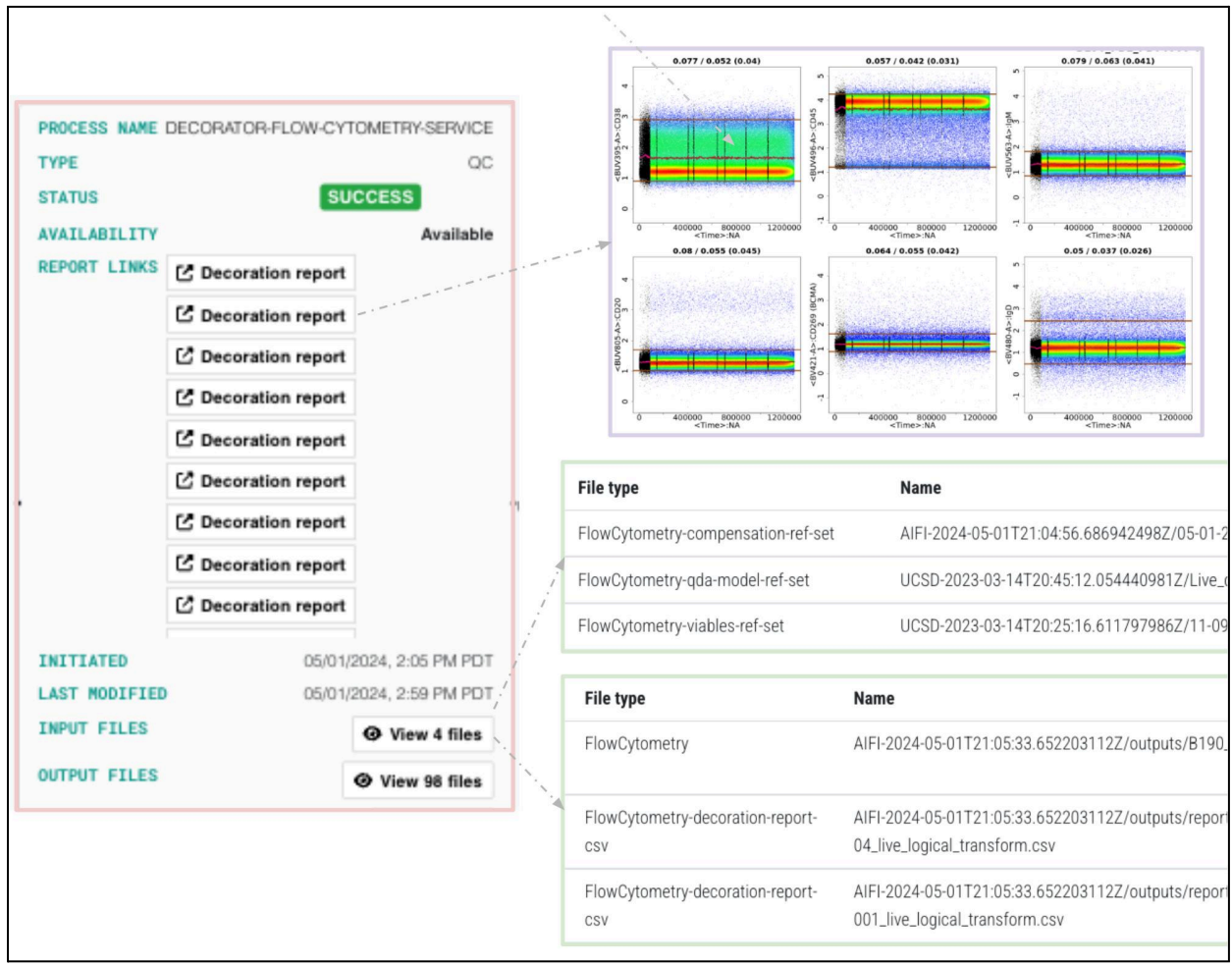

Figure 3. An example of structural metadata for a transformation step.

In complex analysis pipelines, where the output from one step becomes the input for the next, an intricate network of interconnected processes emerges. An example of this is shown in Figure 4.

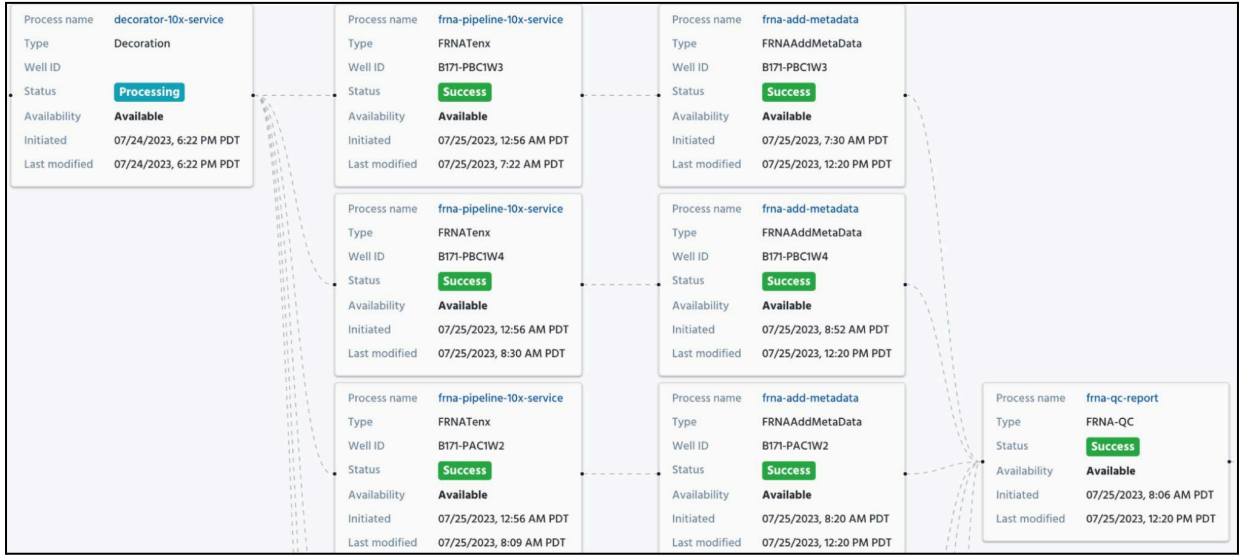

Figure 4. An example of a partial trace of a set of related processes.

# Streamline Administrative Overhead through Automation

Adhering to open science principles by providing insights into data and methods can add to the administrative workload of the researcher. Other open science initiatives have shown that this bureaucratic burden may obstruct adoption and follow-through (Harris, 2018). In the proactive approach, lowering administrative overhead is essential for adoption, as all analyses are tracked whether or not they lead to publication. Automating as much of the tracing process as possible is critical to lowering overhead and encouraging researchers to adopt the approach. In the computational framework described here, researchers' actions, such as data access and running computations, are mediated through an abstraction layer. This layer enables machine processing for tracking their actions. Analysts' manual input is limited to key moments in the process, such as selecting specific data subsets for analysis or defining parameters for generating new results.

Proactive, largely automated tracing not only reduces overhead during analysis, but also reduces the burden of support after publication. Research on the analysis reproducibility of published studies that adhere to open science practices shows that in ¼ to nearly ½ of the studies, reproducibility could be achieved only after authors provided more detailed information about their tools and workflow (Hardwicke et al., 2018, 2021). Eliminating or reducing the need to contact authors to request more information will reduce interruptions and should significantly improve the completion rate for replication attempts.

## Advertising Transparency

Traces should be published with a badge verifying that the work is fully transparent. For instance, a trace that provides the entire flow—from raw data acquisition to automated pipeline analysis to the creation of derived results through secondary analysis, including the creation of visualizations to understand those insights—is more transparent than a trace that offers only visualization insights on derived results. Within the reproducibility framework, a *certificate of reproducibility* signifies that published research adheres to transparent workflows, ensuring the replicability of its findings. The ability to earn badges and display them to external scientists has been shown to be effective in promoting open science practices (Kidwell et al., 2016).

To further advertise and reward transparency, the data and the underlying infrastructure provided here should be recognized as its own scholarly output (Lowenberg, 2022). To that end, each release with a certificate of reproducibility is given a digital object identifier (DOI) to enable its direct citation and recognition as a separate body of academic work.

## Executable Tooling

To re-execute data analysis for verification, it is essential that the complete computational infrastructure be made available (Grüning et al., 2018; Heil et al., 2021; Perkel, 2023). In our reproducibility framework, a certificate of reproducibility containing the full trace, a badge and a DOI can be visually inspected by any scientist, meaning that the steps in the analysis can be reviewed. Additionally, an outside scientist can re-execute (portions of) the trace. For instance, they may opt to re-run a secondary analysis that was originally run in a Jupyter Notebook. To avoid variation introduced by the setup of a scientist's local machine, the scientist is provided the full virtual infrastructure containing the original coding environment of that notebook, including the original (virtual) machine, the (scientific) libraries used, the code, and the input data for that analysis. In the same way, if the user wants to re-run an analysis pipeline, this can be done in the provided infrastructure.

## Open Science with Equitable Access

Adherence to open science principles is a central tenet of our reproducibility framework. The chief self-correcting mechanism of science is the scientific community's engagement in debate and evaluation of findings (Oreskes, 2021). To facilitate this debate and easily detect and eliminate implicit assumptions, published research must be accessible and interpretable by a diverse scientific community (Oreskes, 2021; Ross-Hellauer, 2022). Scientific analysis increasingly relies on considerable resources, such as the infrastructure needed to support computational, storage and other resources, and the expertise needed to operate this infrastructure. Given that there are structural inequities between scientists and institutions with regard to these resources (Bezuidenhout & Chakauya, 2018), providing open access to scientific analysis cannot be limited to making the data available but must also include access to the tools and resources needed to verify results (Cole et al., 2022). Our reproducibility framework is designed to provide such equitable access. Any scientist can reproduce and evaluate results directly in the framework using the provided tools, significantly lowering the financial and technological bar to access.

## Case Studies

Next we turn to several studies run and analyzed using Human Immune System Explorer (Meijer, 2022). This platform is built upon the reproducibility principles discussed above. These examples utilize data previously described in (Thomson et al., 2023).

The study "Comparison of Leukapheresis PBMC and Ficoll PBMC scRNA-seq" (https://doi.org/10.57785/96bw-7571) showcases the Human Immune System Explorer's (HISE) provenance tracing functionality, starting with pipeline processing of control data for two scRNA-seq samples, producing labeled scRNA-seq files in HDF5 (.h5) format. The authors then use a Jupyter Notebook to perform a comparative analysis of this data, generating a Data Set that summarizes the results. The [Certificate of Reproducibility](#) demonstrates the use of a standardized pipeline as a primary analysis step, followed by the use of a Jupyter Notebook to perform secondary, cross-sample, analysis.

In the study "Comparison of Pediatric and Older Adult T cells from TEA-seq" (https://doi.org/10.57785/sv3w-w848), TEA-seq assays were used to characterize the T cells from healthy pediatric and young adult samples (N = 4 for each group). The authors employed the reproducibility framework for data analysis, generating a data set and a visualization to showcase final results. A [Certificate of Reproducibility](#) offers complete transparency into the analytical process. As the provenance trace highlights, the authors adopted a modular approach to analysis. This involved breaking down the analysis into discrete steps, each executed within a separate Jupyter Notebook. By persisting all intermediate results, the authors ensured the ability to revisit and review each step independently. This approach not only facilitates exploration of alternative algorithmic approaches during analysis, but also makes the work more transparent and verifiable by the open science community. Since each step is self-contained, it can be easily assessed, modified, and rerun.

# Discussion

The above case studies demonstrate that utilizing an open science computational framework not only enhances comprehension of research but also deepens insights. This framework, along with its guiding principles, can serve as a valuable blueprint for numerous organizations seeking to achieve analysis reproducibility for many organizations. Full support for open science practices, however, requires a multi-pronged approach that goes beyond technological solutions.

## Rewarding Open Science as an Organizational Practice

A key incentive for scientists' active adherence to open science practices is that organizations actively reward and recognize open and responsible research. This includes investing in the resources, infrastructure and training to effectively share data and methods within and outside the research lab (Lowenberg & Puebla, 2022; Martone & Nakamura, 2022b; Waithira et al., 2019). Scientists are more motivated to adopt a computational reproducibility framework based on the guiding principles presented in the current study when research organizations encourage open science and team science practices. Unfortunately, such support appears to be insufficient in many research institutions and academic organizations, and more must be done to encourage it (Cole et al., 2022).

## How Free is FAIR and Equitable Access?

The cost of non-reproducible research is considerable. An estimated $25 billion is spent annually on non-reproducible studies in preclinical research alone (Freedman et al., 2015). In the Freedman et. (2015) study, the authors argue that although implementing measures to improve reproducibility would raise the cost of individual studies, reducing the percentage of non-reproducible studies will yield significant cost savings. An approach as chosen in the current study where a trace is created and made available as analysis unfolds, enables scientific flaws to be caught early, before publication, and limits the release of non-reproducible research.

To effectively support an organization's commitment to transparent research practices, establishing a cost recovery mechanism is essential (Waithira et al., 2019). Securing long-term financial stability through proper funding and resource allocation is essential for both ensuring internal adherence to open science best practices and earning the recognition and trust of external researchers in the data, tools, and technology provided. Our open science framework offers a comprehensive package: access to the underlying framework itself, secure data storage for research materials, and the necessary computational resources for re-running analyses.

Notably, our framework prioritizes equitable access for external researchers. When reproducing analyses, they incur only the cloud infrastructure costs directly associated with their specific storage and computations. If necessary, academic labs can seek grant funding to offset these costs.

## Analysis Reproducibility and Study Reproducibility

Analysis reproducibility is one of the components contributing to full reproducibility. In life sciences, data generation and collection in the wet lab is usually a major focus (Baker, 2016a; Harris, 2018). For example, for our studies in human immunology, major efforts were undertaken to develop standard operating procedures for blood and tissue sample procurement and processing at different geographic sites, including setting tight standards on processing delays (Bumol et al., 2021; Savage et al., 2021). Although a full discussion of best practices in the wet lab is outside the scope of the current study, it should be noted that standardization of data collection in the wet lab enables downstream standardization and greatly facilitates data ingestion and uniform analysis.

Another important facet of open science is promoting collaborative research. A team science approach, emphasizing collaboration from the study's conception and continuing throughout all phases, enhances reproducibility (Begley et al., 2015; Rolland et al., 2020). Team science can be amplified by providing a common platform that fosters transparency by allowing all team members to collaborate on study design and analysis, getting direct access to results, and monitoring progress. Several specialized team science platforms exist, such as the Open Science Framework (Center for Open Science, 2013) and the Human Immune System Explorer (Meijer, 2022), which we discuss here in the context of its analysis reproducibility design principles.

## Reproducibility and Public Trust in Science

Supporting full adoption of the open science principles proposed here is essential to the continued success and recognition of science in all its manifestations, including fundamental and applied science, both publicly and privately funded. Fundamental science, the basic research that many believe is an essential precursor to breakthrough innovation, is sensitive to public funding cuts, as the usefulness of non-applied research is often questioned by the general public (Flexner, 2017). Improving reproducibility would help boost trust in basic research and bolster the case for continued strong public funding. Similarly, withholding the findings of privately funded applied science has hampered and delayed scientific debate on issues ranging from the effects of tobacco smoke to global warming, undermining trust in industry-funded research (Oreskes & Conway, 2011; Supran et al., 2023). In both cases, adherence to open science principles and encouragement of open debate would go a long way toward restoring and strengthening public trust in science.

# Conclusions

Open science principles are fundamental to enhancing research reproducibility and fostering trust in scientific findings. The guiding principles of the reproducibility framework we present here limit scientific flaws, enable easy detection of scientific failure, and can spur a discussion of alternate analysis approaches while providing equitable access for all members of the academic community. Universities, research organizations, and granting agencies play a pivotal role in fostering and rewarding open science and collaborative team science practices in all phases of research. By actively supporting these practices, they can significantly contribute to the advancement of reliable and trustworthy scientific knowledge.

# Acknowledgement

We are thankful to all members of the Allen Institute for Immunology for their support for and dedicated contributions to the Human Immune System Explorer (HISE). This paper and the research behind it would not have been possible without the collaborative team efforts of everyone at the Allen Institute for Immunology. We especially like to thank the High Resolution Translations Immunology team for their contributions to the case study papers. We are indebted to Scott Pegg for the inspiring discussions about reproducibility and data governance, and are grateful to the leadership and support of Rui Costa, President and CEO of the Allen Institute, and Allen Institute founder, Paul G. Allen, for his vision, encouragement, and support.

# Data and code availability

The scientific case studies used to demonstrate Certificates of Reproducibility were generated from data originally described in (Thomson et al., 2023). Human samples for this study were collected following protocols approved by the IRB of Benaroya Research Institute (Older Adult samples) and the IRB of the Children's Hospital of Philadelphia (Pediatric samples). Raw data is deposited in dbGaP for controlled access at dbGaP Study ID phs003400.v1.p1. Processed data is openly available in NCBI GEO at GEO Accession GSE214546. The code used to generate these case studies is available on Github at https://github.com/aifimmunology/certpro-workflow-demos/.